\documentclass[pre,aps,twocolumn,superscriptaddress,longbibliography,10pt]{revtex4-2}
\usepackage{xcolor}
\definecolor{darkblue}{rgb}{0,0,0.6}
\usepackage[colorlinks,linkcolor=darkblue,citecolor=darkblue,urlcolor=darkblue]{hyperref}
\usepackage{amsmath}
\usepackage{amsfonts}
\usepackage{verbatim}
\usepackage{graphicx}
\usepackage{esint}
\usepackage{amssymb}
\usepackage{mathrsfs}
\usepackage{bm}
\usepackage{bbm}
\usepackage{circledsteps}
\usepackage{tikz,standalone}
\usetikzlibrary{matrix,positioning}
\usepackage{physics}
\usepackage{mathtools}
\newcommand{\beq}{\begin{equation}}
\newcommand{\eeq}{\end{equation}}

\usepackage{soul}
\newcommand{\rev}[1]{\textcolor{black}{#1}}

\begin{document}

\title{Identifying the relevant parameters in design strategies for stable glasses}

\author{Leonardo Galliano}

\affiliation{Laboratoire de Physique de l’École normale supérieure ENS, Université PSL, CNRS, Sorbonne Université, Université de Paris, 75005 Paris, France}

\affiliation{Dipartimento di Fisica, Universit\`a di Trieste, Strada Costiera 11, 34151, Trieste, Italy}

\author{Ludovic Berthier}

\email{ludovic.berthier@espci.fr}

\affiliation{Gulliver, CNRS UMR 7083, ESPCI Paris, PSL Research University, 75005 Paris, France}

\date{\today}

\begin{abstract}
A glass is conventionally obtained by cooling a bulk supercooled liquid through its glass transition temperature. The discovery of ultrastable glasses prepared using physical vapor deposition, together with the recent multiplication of numerical algorithms created to increase the stability of glasses, demonstrates the existence of a variety of strategies for designing glasses with different physical properties. This raises a broader question: which parameters most strongly govern the enhancement of glass stability? Existing computational strategies often produce highly stable glasses by optimizing certain physical properties through dynamical changes in particle diameters. We challenge the idea that these physical quantities are causally responsible for glass stability and suggest instead that diameter dynamics is the principal source of enhanced stability. To support our view, we introduce computational methods to optimize physical quantities without changing the particle diameters. Using the examples of enhanced hyperuniformity at large scale and local ordering at small scale, we design glass configurations with highly optimized values compared to bulk equilibrium states. However, these glasses do not show enhanced stability. The proposed physical quantities are correlated with glass stability, but are not causally responsible for ultrastability. These findings indicate that design rules for stable glasses should be reinterpreted in terms of the dynamical processes that generate stability, rather than the optimized physical quantities they target.
\end{abstract}

\maketitle

\section{Introduction}

Glasses are amorphous solids that form when a liquid is cooled sufficiently rapidly to avoid crystallization~\cite{berthier2016facets}. They display mechanical and dynamical properties that depend sensitively on their preparation history, such as, for instance, the cooling rate. As a result, glass formation is not a uniquely-defined process, but should rather be seen as a broad class of non-equilibrium pathways leading to arrested amorphous states, with varying degrees of stability.

By far, the most common route to glass formation consists in cooling a supercooled liquid through its glass transition temperature~\cite{angell1995formation,royall2018race}. In this case, the system falls out of equilibrium when the structural relaxation time becomes larger than the experimental or numerical observation time. The resulting glass is therefore kinetically trapped, and its properties depend mainly on the rate of cooling. In particular, slower cooling generally leads to more stable glasses that access deeper states in the potential energy landscape. It is however difficult to vary the preparation timescale over a sufficiently broad range to drastically change the physical properties in a controlled way. 

In recent years, it has been shown that significantly more stable glassy states can be obtained using alternative design strategies. Experimental advances, most notably physical vapor deposition~\cite{swallen2007organic,ediger2017highly,rodriguez2022ultrastable}, have demonstrated the possibility to produce glasses with properties comparable to those of extremely slowly-aged systems. In parallel, numerical developments have introduced a range of computational design strategies that similarly aim at increasing glass stability by modifying the dynamical rules or sampling procedures used during preparation~\cite{berthier2023modern,barrat2023computer,ninarello2026}. These recent developments indicate that glass stability can in fact be substantially enhanced when the system is adequately guided toward deeper regions of its energy landscape.

A wide variety of design strategies for stable glasses has emerged, often combining several modifications of microscopic dynamics or sampling protocols~\cite{zhang2016perfect,berthier2017origin,brito2018theory,yana2021towards,varda2002transient,dale2022hyperuniform,fan2024ideal,wang2025hyperuniform,leoni2025generating,bolton2026ideal,kapteijns2019fast}. These design strategies typically achieve their goal of producing more stable glasses, but they do so through complex and sometimes intertwined procedures. As a consequence, it is difficult to identify which elements of these design strategies are most relevant for achieving increased glass stability. In particular, several physical quantities have been systematically optimized across different strategies~\cite{zhang2016perfect,yana2021towards,dale2022hyperuniform,fan2024ideal,wang2025hyperuniform,leoni2025generating,brito2018theory,kapteijns2019fast}, leading to the hypothesis that they may play a central role in controlling stability.

These results raise a more general question about the identification of the relevant parameters underlying design strategies for stable glasses~\cite{berthier2026designing}. In particular, we note that while many studies report correlations between specific observables and glass stability, it is not clear whether these quantities act as causal control parameters or whether they simply reflect the outcome of the dynamical processes used to generate stable configurations. It is for instance puzzling that different design strategies may lead to similar improvements in stability while targeting quite distinct sets of observables. This could suggest that these observables either do not constitute independent control parameters, or, even more radically, that they do not control glass stability at all.

In this work, we revisit computational design strategies for stable glasses to disentangle the role of dynamical modifications and optimized properties, which appear as common ingredients across several strategies. To this end, we introduce controlled computational methods that allow us to optimize selected physical quantities while keeping the underlying particle identities and particle diameters fixed. This enables us to separate the effect of optimized observables from that of the dynamical processes used to generate them. Using two representative examples, involving enhanced hyperuniformity at large length scales~\cite{torquato2018hyperuniform} and increased local structural order at small scales~\cite{tong2018revealing}, we construct glass configurations in which these quantities are strongly optimized compared to equilibrium bulk states. However, despite these significant changes in structural observables, we do not observe a corresponding increase in glass stability. These results indicate that stable glasses have optimized physical quantities, but the opposite is not true: configurations with enhanced properties do not necessarily display ultrastability. These quantities are thus correlated with glass stability but are not causally responsible for it. 

\rev{We interpret these results to mean that the applied optimizations merely reorganize the local structure within the original glassy basin. While the arrested states possess sufficient internal degrees of freedom to allow structural properties to be optimized, these local rearrangements are not enough to drive the system toward deeper, more stable regions of the energy landscape.}

Overall, our results suggest that the effectiveness of several design strategies for stable glasses arises from the underlying dynamical processes used to generate low-energy configurations rather than from the choice of a specific observable chosen to be optimized. This is a reasonable conclusion: provided that these quantities are sufficiently correlated to the potential energy, optimizing them has an effect similar to optimizing the energy itself. This conclusion points toward a unified interpretation of design strategies for stable glasses in terms of their underlying dynamical mechanisms rather than in terms of the particular physical quantities they aim to optimize.

In Sec.~\ref{sec:model} we define the chosen numerical model and study its bulk properties. 
In Sec.~\ref{sec:hyperuniformity}, we devise a numerical strategy to enhance hyperuniformity of density fluctuations at large scale. In Sec.~\ref{sec:theta} we use biased Monte Carlo simulations to enhance a local packing property. We summarize and discuss our results in Sec.~\ref{sec:conclusion}. 

\section{Numerical model and its bulk properties}

\label{sec:model}

\subsection{Glass model and local Monte Carlo simulations}

We study a two-dimensional model of hard disks using a binary distribution of diameters with diameter ratio 1:1.4 and composition 65:35. We set the diameter of the smaller particles as our unit of length. This model is convenient for our purposes because it is an excellent glass-former (it does not crystallize easily), it is easy to simulate using conventional Monte Carlo methods~\cite{frenkelUnderstandingMolecularSimulation2023}, and the aspect ratio is such that swap Monte Carlo techniques~\cite{grigera2001fast,berthier2016equilibrium,ninarello2017models} cannot be used for this model (swap moves are essentially all rejected). There exists therefore no known method to provide highly stable configurations for this model, and only conventional numerical tools are available. It is thus a good model to benchmark non-conventional design rules for stable glasses against conventional local Monte Carlo methods.  

We use local Monte Carlo translational moves to study the equilibrium properties of the model~\cite{berthier2007monte}. In this approach particles are chosen sequentially and a small translation is proposed within a square box of linear size $\delta=0.1$. These moves are accepted if they create no overlap. This local Monte Carlo algorithm satisfies detailed balance and samples the equilibrium Boltzmann distribution. We use $NPT$ Monte Carlo simulations using $N=1000$ particles in a box of linear size $L$ that varies dynamically by small amounts to keep the imposed pressure $P$ to the desired constant, again using volume changes that satisfy detailed balance~\cite{frenkelUnderstandingMolecularSimulation2023}.   

At a given imposed pressure $P$, we can measure the packing fraction $\phi = \pi \overline{\sigma^2} N / (4 L^2)$, where the overline is an average over the size distribution. The packing fraction is proportional to the number density $\rho = N / L^2$, and it is dimensionless. Similarly, it is convenient to define the compressibility factor (or reduced pressure) $Z = P / (\rho k_B T)$, because it is proportional to the imposed pressure $P$ but is also dimensionless. The relation between $Z$ and $\phi$ defines the pressure-density equation of state, $Z=Z(\phi)$ using the dimensionless variables that are most appropriate for hard disks. Because $P$ and $T$ only enter as a ratio $P/T$ for hard particles, changing $P$ at constant $T$ or changing $T$ at constant $P$ are equivalent procedures, so that $1/Z \sim T/P$ is equivalent to temperature for a system with soft interactions~\cite{berthier2009glass}. For each $Z$ value, we perform Monte Carlo simulations until thermal equilibrium is reached. We then measure static and dynamic properties, as reported below. For dynamic properties, we define the time unit as $N$ attempts to perform a local Monte Carlo move. 

\subsection{Equilibrium slow dynamics}

We start by investigating the equilibrium dynamics of the system to determine over which density regime it displays an interesting glassy regime. To this end, we determine the self-intermediate scattering function
\begin{equation}
    F_s(q,t) = \frac{1}{N} \langle \sum_i e^{i {\mathbf q} \cdot ({\mathbf r}_i(t) - {\mathbf r}_i(0) ) } \rangle,
\end{equation}
where $q= |\mathbf{q}| \approx 6.1$ is chosen near the first peak of the static structure factor, $\mathbf{r}_i(t)$ represents the position of particle $i$ at time $t$, and the brackets represent an ensemble average. The sum is over all particles, independently of their size. 

\begin{figure}
\includegraphics[width=\columnwidth,clip=true]{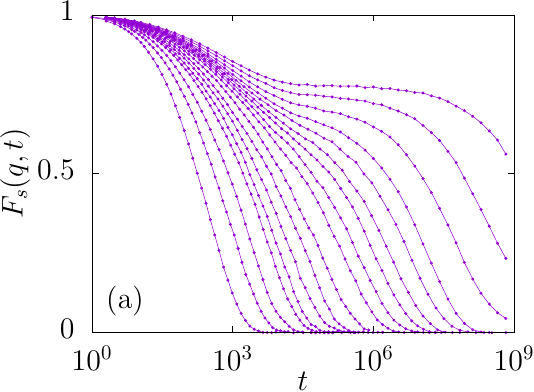} 
\includegraphics[width=\columnwidth,clip=true]{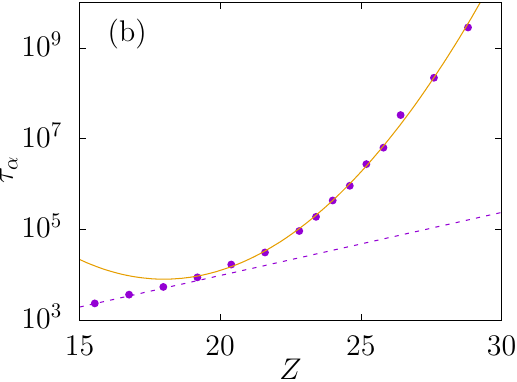} 
\caption{Equilibrium slow dynamics of the bulk model. 
(a) Self-intermediate scattering function $F_s(q,t)$ at different values of the reduced pressure $Z$ measured from $NPT$ simulations, from $Z=15$ to \rev{$Z=28.8$} (from left to right). 
(b) Pressure evolution of the relaxation time $\tau_\alpha$ as a function of $Z \sim P/T$, in a plot similar to the conventional Arrhenius plot for systems with soft interactions. The dashed line shows Arrhenius behavior at low pressures (high temperatures), locating the onset of slow dynamics near $Z \approx 20$. The full line is a fit to a parabolic law. Both quantities display typical signatures of glassy dynamics as $Z$ increases ($T$ decreases).}  
\label{fig:fsqt}
\end{figure}

We show in Fig.~\ref{fig:fsqt}(a) the time evolution of $F_s(q,t)$ for a range of $Z$ values from $Z \approx 15$ to $Z \approx 29$. Above $Z \approx 20$, which corresponds to the onset of slow dynamics, we observe the well-known two-step decay of time correlation functions, with a slow decay at long times that is very sensitive to the pressure $Z$. Above $Z\approx29$, the relaxation is too slow, equilibration and equilibrium sampling become difficult within a reasonable computer time.

We extract a structural relaxation time $\tau_\alpha$ from the time decay of the self-intermediate scattering function, using the definition $F_s(q, t=\tau_\alpha) = 1/e$. We represent its evolution with pressure $Z$ in Fig.~\ref{fig:fsqt}(b). Because $Z \sim P/T$ this representation is analogous to an Arrhenius plot for systems with soft pair interactions~\cite{berthier2009glass}. We observe that the onset for slow dynamics is near $Z \approx 20$, above which the data depart from a low-pressure Arrhenius behavior, while equilibration becomes challenging above $Z \approx 29$. The narrow range of $Z = 20-29$ is the relevant regime where the system painfully tries to access deeper states in the potential energy landscape, and the structural relaxation time grows by a factor $\approx 10^5$ over this relevant range of pressures. This dynamic range directly dictates the range of glass stabilities that can be obtained using conventional computation methods for a simple glass-forming model. These different stabilities will serve as a benchmark to assess the effect of optimizing structural properties.  We also add a parabolic fit~\cite{elmatad2009corresponding} description of the data in the slow relaxation regime. This shows that the model displays, in this dynamic range, the behavior of fragile glasses. Below, we will also use this parabolic fit to estimate timescales for larger pressures, where direct simulations are no longer possible~\cite{berthier2020how}.  

\rev{For the largest pressure, $Z=28.8$, simulations were equilibrated for $10^{10}$ steps prior to measuring the intermediate scattering function, corresponding to roughly 10 times the measured relaxation time ($\tau_\alpha \approx 2 \times 10^9$). The time correlation was further averaged over subsequent production runs of the same duration without revealing any drift, confirming that the system is still close to equilibrium, although admittedly we are on the border of the equilibration window.}

\subsection{Some static bulk properties}

In the course of equilibrium simulations, we measure the static structure factors, the isothermal compressibility, and the local packing order parameter $\Theta$ that will be used later as targeted properties for designing non-equilibrium glassy states.

To detect hyperuniformity~\cite{torquato2018hyperuniform} at large length scales in a binary mixture, it is convenient to measure the $q$-dependent isothermal compressibility, $\chi(q)$~\cite{berthier2011suppressed}. For a bidisperse mixture with species $\mathrm{A}$ and $\mathrm{B}$ at concentrations $x_\mathrm{A}$ and $x_\mathrm{B}$, the compressibility is defined as~\cite{bhatia1970structural}
\beq
\chi(q)=\frac{S_{\mathrm{AA}}(q)S_{\mathrm{BB}}(q)-S_{\mathrm{AB}}^2(q)}{x_{\mathrm{A}}^2S_{\mathrm{BB}}(q)+x_{\mathrm{B}}^2S_{\mathrm{AA}}(q)-2x_{\text A}x_{\mathrm{B}}S_{\mathrm{AB}}(q)},
\eeq
where the partial structure factors are defined as:
\beq
S_{\alpha\beta}(q)=\frac{1}{N}\sum_{i \in \alpha} \sum_{j \in \beta} \left< \mathrm e^{ i  {\mathbf q} \cdot \left( {\mathbf r}_i- {\mathbf r}_j \right)}\right> ,
\eeq
with $\alpha, \beta = \mathrm{A}, \mathrm{B}$. In a hyperuniform system, $\chi(q)$ vanishes as a power law $\chi(q) \sim q^\alpha$ with $\alpha > 0$ as $q \to 0$, indicating suppression of density fluctuations at large scale. In contrast, in equilibrium fluids and conventional glasses obtained by cooling, $\chi(q)$ approaches a finite constant as $q \to 0$. This constant is directly related to the isothermal compressibility in fluid states, and indirectly linked to bulk and shear moduli in glassy solid states~\cite{ikeda2015thermal}. 

\begin{figure}
\includegraphics[width=\columnwidth,clip=true]{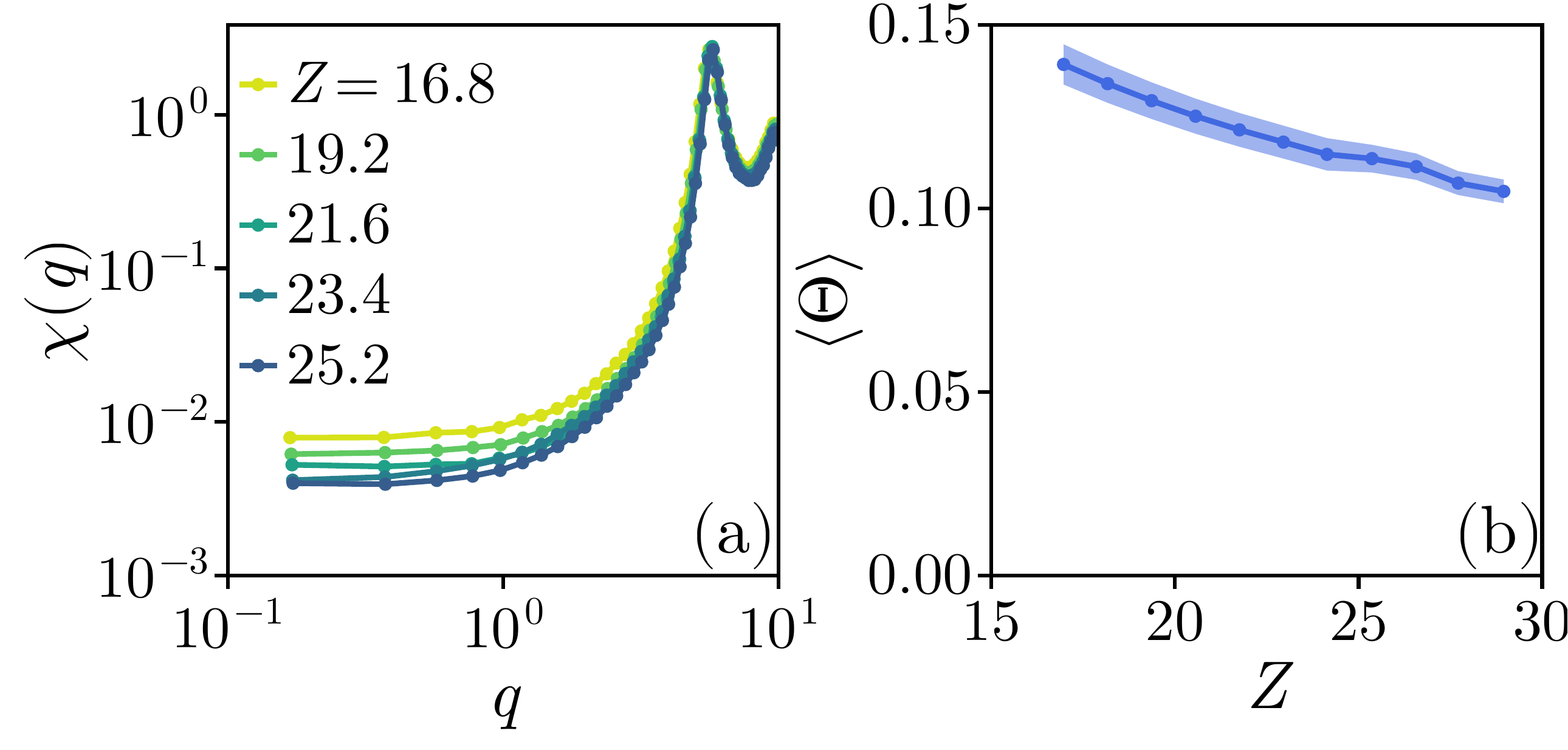} 
\caption{Static bulk properties in thermal equilibrium. 
(a) $q$-dependent isothermal compressibility $\chi(q)$ at different reduced pressures $Z$, in thermal equilibrium. The log-log representation emphasizes the fast approach to a plateau at low $q$, that decreases weakly with increasing $Z$ in a system of $N=1000$ particles. 
(b) Average local order parameter $\left<\Theta\right>$ as a a function of $Z$ in equilibrium. The shaded area indicates the standard deviation of the fluctuations of $\Theta$ in systems of $N=200$ particles.}
\label{fig:statics}
\end{figure}

We show in Fig.~\ref{fig:statics}(a) the wave vector dependence of $\chi(q)$ for a system of $N=1000$ particles measured in equilibrium conditions for different compressibility factors $Z$. The $q$-dependence is as expected for equilibrium fluid states. The local structure of the fluid yields a peak near $q \approx 6.1$ that essentially reflects the interparticle distance in the dense fluid. At low $q$ the compressibility converges to a constant, that evolves weakly over the range $\chi (q \to 0) \approx 0.006$ near the onset, to $\chi (q \to 0) \approx 0.003$ close to the numerical glass transition at larger $Z$. Clearly, there is no observed tendency towards hyperuniform behavior, and the compressibility converges quickly to its low-$q$ limit when $q$ decreases below the first diffraction peak. The evolution with pressure is also quite modest, which is consistent with the conventional observation that the structure of supercooled liquids, at two-point density correlation level, evolves very little as the glass transition is approached~\cite{berthier2011theoretical}.  

The local packing order parameter, usually denoted $\Theta$, was introduced in Ref.~\cite{tong2018revealing}. It quantifies local deviations from an ideal, perfectly packed configuration and is defined as follows. For each particle $n \in [1, N]$, we first construct its Voronoi tessellation to determine its set of neighbors. For any pair of neighbors $(i,j)$ of $n$ that are also neighbors of each other, the triplet $(n,i,j)$ forms a Delaunay triangle. Let $\theta^1_{ij}$ be the angle at vertex $n$ between vectors ${\mathbf r}_{ni}$ and ${\mathbf r}_{nj}$ in the simulated configuration. We then determine $\theta^2_{ij}$ as the corresponding angle in an idealized packing configuration where the three particles $(n,i,j)$ are just touching. The difference between $\theta^1_{ij}$ and $\theta^2_{ij}$ quantifies the local deviation from a perfect packing of disks. For each particle $n$ we then define 
\beq
\Theta_n = \frac{1}{N_n} \sum_{\left<i,j\right>} \left|\theta^1_{ij}-\theta^2_{ij}\right|,
\eeq
where the sum runs over the $N_n$ Delaunay triangles defined above that surround particle $n$. The global packing order parameter for a given configuration containing $N$ particles is then obtained as an average over particles, $\Theta = N^{-1} \sum_n \Theta_n$, and its ensemble average $\langle \Theta \rangle$ then defines the average order parameter.  

We have measured the evolution of $\langle \Theta \rangle$ with pressure, as shown in Fig.~\ref{fig:statics}(b), as well as its standard deviation measured in systems composed of $N=200$ particles. We find that $\left<\Theta\right>$ slightly decreases smoothly with increasing $Z$, indicating that the local order increases weakly with compression in this regime. In particular, while the structural relaxation time $\tau_\alpha$ grows by roughly five orders of magnitude over this range, see Fig.~\ref{fig:fsqt}(b), $\left<\Theta\right>$ decreases by about $20\%$. This is again consistent with the idea that local structure evolves weakly in the supercooled liquid regime. 

The measurements in Fig.~\ref{fig:statics} show that configurations that are equilibrated at larger pressures display smaller density fluctuations at large scales and better packing at the local scale. These two static quantities, that are very different in nature, are therefore clearly both correlated with the stability of the system.  

\subsection{Fluid and glass equations of state}

\label{sec:glassEOS}

\begin{figure}
\includegraphics[width=\columnwidth,clip=true]{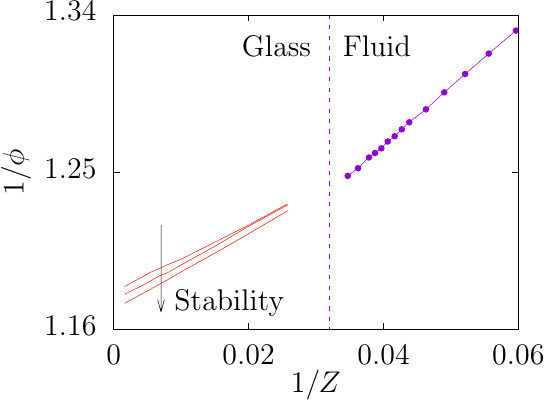} 
\caption{Equation of state in the $1/\phi$ (free volume) versus $1/Z \sim T/P$ representation analogous to energy (or free volume) versus temperature evolution for soft potentials. The equilibrium equation of state for the fluid ends near the computational glass transition (vertical dashed line). In the glass phase at larger pressure, the equation of state becomes history-dependent. We construct three different glass histories (fast compression from fluid states at $Z_{\rm init}=16.8$, $22.8$ and $28.8$) leading to three glasses of different stabilities from poorly annealed to very stable (top to bottom).}   
\label{fig:eos}
\end{figure}

In the course of the equilibrium simulations, we also measure the equation of state of the system $Z=Z(\phi)$ for the fluid. It is convenient to represent the evolution of the free volume $1/\phi$ versus $1/Z \sim T/P$, because this graph is then fully analogous to the evolution of potential energy versus temperature in systems with soft interactions. It also resembles the typical graph depicting the temperature evolution of free volume in glass-formers across their glass transitions, which usually serves as an introduction to the glass transition problem~\cite{berthier2016facets}. Results for the equilibrium fluid are shown in Fig.~\ref{fig:eos}, where we observe a smooth decrease of the free volume when $Z$ is increased ($T$ is decreased). These equilibrium measurements stop when the numerical glass transition is approached near $Z\approx 29$ (indicated with a vertical dashed line). 

When pressure is increased beyond $Z \approx 29$ it is no longer possible to determine the equilibrium equation of state and the system enters the glass region. In that case, the equation of state explicitly depends on the preparation of the glass, with the expectation that a slower annealing of the system should lead to deeper regions of the potential energy landscape, where glasses will be more stable and denser~\cite{fullerton2017density}. 

To illustrate this point we design three preparation protocols yielding glasses with three different  stabilities, as shown in Fig.~\ref{fig:eos}. We first prepare a fluid configuration at a given pressure $Z_{\rm init}$ in the fluid phase. We suddenly compress this fluid configuration to a very large pressure, $Z=500$, where we let it age for $t=10^5$ Monte Carlo steps. At such a large $Z$, particle motion is rapidly arrested and the glass state is prepared. We then slowly decompress these glass configurations, and record the equation of state $Z(\phi)$ of that glass during the decompression. We checked that no particle rearrangement occurs during the decompression. Therefore this protocol truly follows the non-equilibrium equation of state of a structurally arrested glass state. We can then easily vary the stability of the glass by changing the value $Z_{\rm init}$ of the initial fluid configuration. We expect that lower $Z_{\rm init}$ values will yield less stable glasses. 

We report the measured glass equations of state for three values of $Z_{\rm init}$ in Fig.~\ref{fig:eos}. These data provide the well-known behavior that more stable glasses, originating from larger $Z_{\rm init}$, are also denser. The chosen values of $Z_{\rm init}$ correspond to fluid states with wildly distinct equilibrium relaxation times that vary by about 6 orders of magnitude. Therefore the relatively modest change in the three equations of state in Fig.~\ref{fig:eos} actually represent a rather large change in the preparation protocol, from a very poorly annealed glass to the largest stability that can be achieved when using conventional numerical algorithms. Given the available computer time, it would be very difficult to improve on the glass stability using local Monte Carlo simulations. These data therefore serve as a benchmark to test the gain in stability offered by alternative computational routes. 

\rev{The glass equation of state $Z=Z(\phi)$, which we represent in the graphical form $1/\phi$ versus $1/Z$, allows us to directly visualize and quantify how deep in the energy landscape different glasses lie. Below, we compare these glass equation of states to establish whether different glasses access deeper valleys in the free energy landscape. This can be seen as a metric for thermodynamic stability, which is conceptually different from the kinetic stability also used in the literature~\cite{swallen2007organic}. The latter concept relies on dynamic measurements and aims to describe the kinetics of transformation of stable glasses back into the supercooled liquid state, following either a gradual or sudden change in a control parameter (such as heating). Evidence so far suggests that thermodynamic and kinetic stabilities evolve together~\cite{ediger2017highly,fullerton2017density}.}

\section{Optimizing hyperuniformity at large length scales}

\label{sec:hyperuniformity}

\subsection{Non-equilibrium algorithm to enhance hyperuniformity}

We observed in Fig.~\ref{fig:statics}(a) that configurations generated using an equilibrium Monte Carlo algorithm do not display hyperuniform density fluctuations at large length scales. To produce hyperuniform configurations, we thus need to design glasses using non-equilibrium design rules~\cite{torquato2018hyperuniform,zhang2016perfect}. 

In several previous works~\cite{yana2021towards,dale2022hyperuniform,wang2025hyperuniform}, hyperuniform configurations were obtained by dynamically modifying the particle diameters such that each particle occupies the same area fraction locally, thereby suppressing density fluctuations. The dynamic evolution of particle diameters, in this approach, is not constrained by rules of thermal equilibrium and it indeed drives the system out of equilibrium. Since changing the particle diameters may produce overlaps, these are subsequently removed by another non-equilibrium procedure (such as minimizing the energy with a soft repulsive potential). Performing multiple cycles of diameter dynamics and energy minimization then leads to glassy states with both a strong degree of hyperuniformity and a large kinetic stability~\cite{yana2021towards,dale2022hyperuniform,wang2025hyperuniform}. There are three important features in these non-equilibrium algorithms: (i) diameters and positions are free degrees of freedom, (ii) the particle size distribution is not conserved by the algorithm, (iii) the dynamics happens far from equilibrium. 

Our goal is to devise an algorithm that produces a strong degree of hyperuniformity without altering the particle diameters, thus also preserving the original particle size distribution. To this end, we introduce a non-equilibrium dynamic algorithm that is known to enhance hyperuniformity in dense configuration of hard disks. This non-equilibrium dynamics is borrowed from the field of non-equilibrium phase transitions in driven particle models~\cite{corte2008random}. In particular, the dynamics known as conserved biased random organization is known to drive particle systems to non-equilibrium dynamic steady states where density fluctuations are strongly suppressed at large length scales~\cite{hexner2017noise,hexner2017enhanced,wilken2021random,galliano2023two,galliano2026glass}.

Briefly, the dynamics evolves particle positions in discrete time steps. At each time step, pairs of overlapping particles are given a random repulsive displacement of equal amplitude along the line connecting their centers~\cite{hexner2017noise}. The magnitude of the displacement is uniformly drawn in the interval $[0, \epsilon]$, and multiple overlaps contribute additively. The central point of these dynamics is that the center of mass of the system is conserved at each pairwise collision, and this conservation law was shown to produce hyperuniform structures at large scale~\cite{hexner2017noise}. There are two control parameters in this system: the area fraction $\phi$ of the system, and the maximal amplitude $\epsilon$ of the random jumps. 

To produce hyperuniform hard disk glasses via random organization, we employ the following protocol. We first prepare equilibrium configurations with $N=1000$ particles using local Monte Carlo dynamics at initial pressure $Z_{\rm init}$ and corresponding area fraction $\phi_{\rm init}$. We then switch to the biased random organization dynamics. Since there is no overlap in the original hard disk system and the algorithm needs at least one overlap to start, we apply to each particle a random isotropic displacement with uniformly distributed direction and magnitude in $[0,0.22]$. These disturbed configurations are then evolved using the biased random organization dynamics for \rev{$5\times10^8$} time steps at $\epsilon\in[0.18,0.23]$, depending on the initial packing fraction, until a hyperuniform steady state is reached. Finally, to recover a correct hard disk configuration we switch to a very small value of $\epsilon = 0.02$, that rapidly drives the system back to a configuration with no overlap via a succession of small repulsive displacements. At the end of this process, we thus have hard disk configurations with no overlap and suppressed density fluctuations at large length scales, and these have been obtained without any dynamic coupling between diameters and positions. 

\begin{figure}
\includegraphics[width=\columnwidth,clip=true]{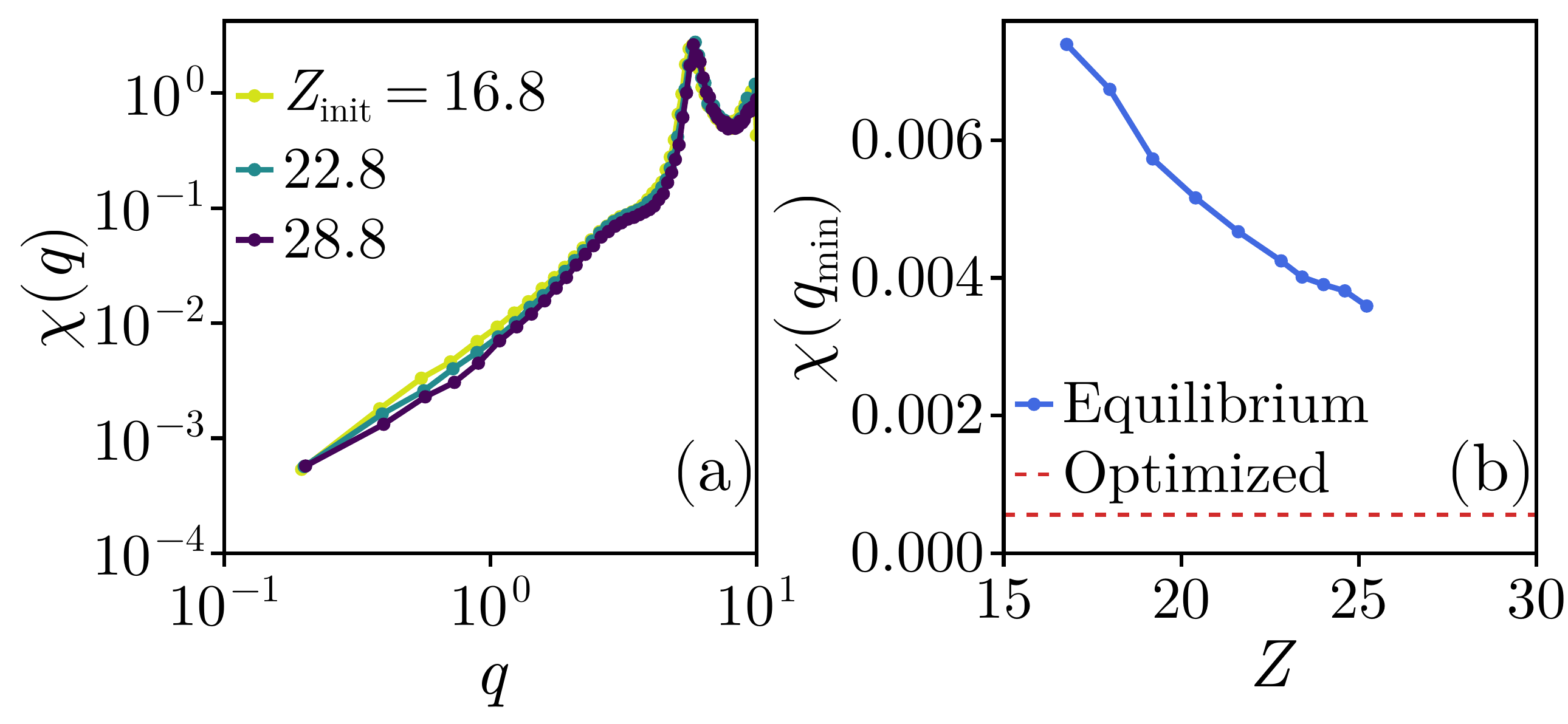} 
\caption{Optimizing hyperuniformity at large length scales.
(a) Isothermal compressibility $\chi(q)$ measured in hard disk configurations prepared at equilibrium at pressure $Z_{\rm init}$ and then driven out of equilibrium using the biased random organization dynamics. A clear hyperuniform behavior $\chi(q) \sim q^2$ is observed at low $q$. 
(b) Comparison of the value of $\chi(q_{\min})$ for optimized (red dashed line) and equilibrium (blue symbols) configurations. The optimized configurations display a degree of hyperuniformity that would only be achieved for $Z \approx 36$ in equilibrium conditions, a regime that is not accessible using conventional numerical methods.}   
\label{fig:chiq}
\end{figure}

We repeated this approach using three values for $Z_{\rm init}$. We show in Fig.~\ref{fig:chiq}(a) the isothermal compressibility $\chi(q)$ measured at the end of the protocol described above. We emphasize that because the final relaxation step to remove overlaps is very quick and involves a small number of local displacements, it does not alter the hyperuniformity of the system, which is inherently a large-scale property. Compared to the equilibrium results shown in Fig.~\ref{fig:statics}(a) the compressibility is now strongly suppressed at low-$q$ and no finite plateau is reached over the range of wavevectors shown in Fig.~\ref{fig:chiq}(a). As discussed in the context of random organization models~\cite{hexner2017noise}, the conserved dynamics typically leads to a quadratic behavior, $\chi(q) \sim q^2$, at least when $\phi$ is not too large~\cite{galliano2026glass}. Compared to the equilibrium configurations produced using local Monte Carlo dynamics with detailed balance, these configurations display enhanced hyperuniformity at large scales. 

To quantify this effect, we introduce $q_{\min}= 2 \pi / L$ which is the smallest accessible wavevectors in the simulations (due to periodic boundary conditions and the finite linear size of the system) and use the value of $\chi(q_{\min})$ as a practical measure of hyperuniformity. The data in Fig.~\ref{fig:chiq}(a) suggest that the value  $\chi(q_{\min}) \approx 5 \times 10^{-4}$ is reached, which is far below the equilibrium values displayed in Fig.~\ref{fig:statics}(a). 

In Fig.~\ref{fig:chiq}(b) we compare equilibrium and non-equilibrium values for $\chi(q_{\min})$. A naive extrapolation of the equilibrium data suggests that reaching comparable $\chi(q_{\min})$ values through conventional $NPT$ simulations would require equilibrium configurations at reduced pressure $Z \approx 36$, corresponding to equilibration times that are many orders of magnitude larger than those we can simulate. Using the parabolic fit to the equilibrium data, we estimate that $\tau_\alpha \sim 10^{19}$ for $Z=36$. Independently of the quality of the extrapolation, such low value of $\chi(q_{\rm min}$ corresponds to highly stable glasses. 

We conclude that by using non-equilibrium random organization dynamics, we can suppress density fluctuations at large length scales efficiently while preserving the hard disk nature of the pair potential and leaving the particle size distribution unaffected. Crucially, this procedure does not explicitly use swap moves or any other known acceleration trick. We are thus in a good position to test whether optimizing hyperuniform behavior at large scale alone can improve the stability of the glass configurations.  

\subsection{Does hyperuniformity drive glass stability?}

\rev{To test the thermodynamic stability} of the hyperuniform glass configurations obtained by non-equilibrium random organization dynamics we repeat the measurement of the glass equations of state, as described in Sec.~\ref{sec:glassEOS} above. 

We quantitatively compare two families of approaches. On the one hand, we can use the equilibrium (non-hyperuniform) hard disk configurations at $Z_{\rm init}$ to produce conventional glasses obtained by rapid compressions to large pressure, as already done in Sec.~\ref{sec:glassEOS}. On the other hand, we can also compress the optimized hyperuniform configurations to large pressures, and measure their glass equations of state during the same decompression protocols.  

\begin{figure}
\includegraphics[width=\columnwidth]{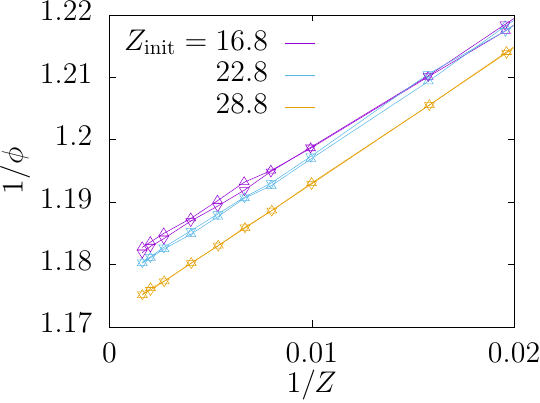} 
\caption{Glass equations of state for glasses prepared using either conventional Monte Carlo methods (upwards triangles) or hyperuniform glasses prepared using random organization dynamics (downwards triangles), for three values of $Z_{\rm init}$. Hyperuniform and conventional glasses are equally stable.}
\label{fig:EOSHU}
\end{figure}

Since we have three values of $Z_{\rm init}$ and two families of glasses (conventional and hyperuniform), we report six glass equations of state in Fig.~\ref{fig:EOSHU}. Each color represents a given value of $Z_{\rm init}$ while different symbols are used to distinguish conventional from hyperuniform glasses. Clearly, the data in Fig.~\ref{fig:EOSHU} are grouped by colors. This implies that glass stability is uniquely controlled by $Z_{\rm init}$, independently of the degree of hyperuniformity. In other words, optimizing hyperuniformity to a degree that can only be obtained in very stable glasses has not influenced the initial stability at all. The logical conclusion is that density fluctuations are suppressed in highly stable glasses, but glasses with suppressed density fluctuations are not necessarily stable. Hyperuniformity and stability are correlated physical properties, but hyperuniformity is not causally responsible for enhanced stability. 

\rev{We have performed some tests to assess whether thermodynamic and kinetic stabilities lead to similar conclusions. A clean protocol to quantify kinetic stability consists of a sudden change of the pressure from a value deep in the glass to a final pressure $Z$ that corresponds to the fluid state. In such a protocol, the packing fraction evolves in two steps: a fast expansion of the glass, followed by a much slower transformation of the glass structure back into the equilibrium fluid at the imposed pressure. The timescale of the second process, when expressed in units of $\tau_\alpha(Z)$, defines the stability ratio ${\cal S}$~\cite{sepulveda2014role,whitaker2015kinetic,fullerton2017density}, which is a useful metric for kinetic stability. We found that the most stable glasses we can achieve using slow cooling have ${\cal S} \approx 10^2$ (for $Z_{\rm init}=28.8$~\cite{fullerton2017density}). Most importantly, we also find that ${\cal S}$ is not sensitive to the degree of hyperuniformity, thus directly showing that thermodynamic and kinetic stabilities lead to similar conclusions.}

Earlier observations of hyperuniform stable glasses resulted from two ingredients~\cite{yana2021towards,dale2022hyperuniform,wang2025hyperuniform}: a coupled position-diameter dynamics and non-equilibrium rules that optimize hyperuniformity. Our interpretation is that it is the coupled position-diameter dynamics that is causally responsible for enhanced stability, rather than the enhanced hyperuniformity. We discuss this interpretation further in the discussion part in Sec.~\ref{sec:conclusion}. 

\section{Optimizing a local packing order parameter}

\label{sec:theta}

\subsection{Biased Monte Carlo algorithm}

We showed in Fig.~\ref{fig:statics}(b) the variation of the local order parameter $\langle \Theta \rangle$ upon approaching the computer glass transition. Our goal in this section is to introduce a numerical algorithm that produces configurations with a much lower value of $\langle \Theta \rangle$, which would be typical of much more deeply supercooled states, possibly yielding ultrastability. 

To this end, the obvious option is to enhance equilibration dramatically to reach even lower values of $\langle \Theta \rangle$. For the present system, however, we know of no algorithm to achieve significantly faster equilibration. Therefore, the only alternative is to explore non-equilibrium strategies, which bypass the equilibrium Boltzmann distribution to directly generate optimized configurations via a non-equilibrium design rule. In this work, we constrain ourselves to methods that do not use diameters as dynamic degrees of freedom and leave the particle size distribution unaffected.  

We follow a general strategy that allows us to target very low values of $\langle \Theta \rangle$. This approach would in fact be applicable to any targeted physical quantity, such as hyperuniformity or global fluctuations of the Virial stress. We perform biased Monte Carlo simulations using a modified Hamiltonian~\cite{frenkelUnderstandingMolecularSimulation2023,landau2021guide}
\beq
H(x) = H_0(x) + \lambda \Theta^2(x),
\label{eq:H}
\eeq
where $H_0(x)$ is the original Hamiltonian defined from the sum of pairwise hard-disk interactions for the $N$-particle configuration $x = \{ \mathbf{r}_1, \cdots \mathbf{r}_N \}$, $\Theta(x)$ is the global value of $\Theta$ in configuration $x$, and $\lambda>0$ controls the strength of the biasing field. Clearly, representative equilibrium configurations of the biased Hamiltonian $H(x)$ are distinct from those of the original Hamiltonian $H_0(x)$. It is also obvious that $H(x)$ should drive the hard disk system out of equilibrium. Finally, the effect of the field $\lambda$ is physically clear: it penalizes configurations with high values of $\Theta$, forcing the system to sample configurations with low values of $\Theta$. 

To sample equilibrium configurations of the Hamiltonian $H(x)$ we introduce the following Monte Carlo procedure. Starting from a configuration $x$, we perform $n$ Monte Carlo time steps using the equilibrium hard-disk Monte Carlo dynamics introduced above. We recall that in this approach, particle displacements are accepted with the Metropolis criterion, ensuring detailed balance with respect to the distribution $P_0(x) \propto \exp \left(-\beta H_0(x)\right)$. The resulting configuration, denoted $x'$, is treated as a proposal in a Metropolis-Hastings algorithm targeting the full distribution $P(x)\propto\exp(-\beta H)$. Because this proposal generation satisfies detailed balance with respect to $P_0(x)$, the standard acceptance probability simplifies to
\beq
\alpha \left(x,x'\right) = \min \left\{1, \exp \left(- \beta \lambda \left(\Theta^2\left(x'\right)-\Theta^2\left(x\right)\right)\right) \right \},
\eeq
to ensure that the Monte Carlo procedure samples $P(x)$. Once the Markov chain reaches steady state, the resulting configurations therefore correspond to equilibrium states of the biased Hamiltonian $H(x)$, which are non-equilibrium states for the original Hamiltonian $H_0$. 

\begin{figure}
\includegraphics[width=\columnwidth,clip=true]{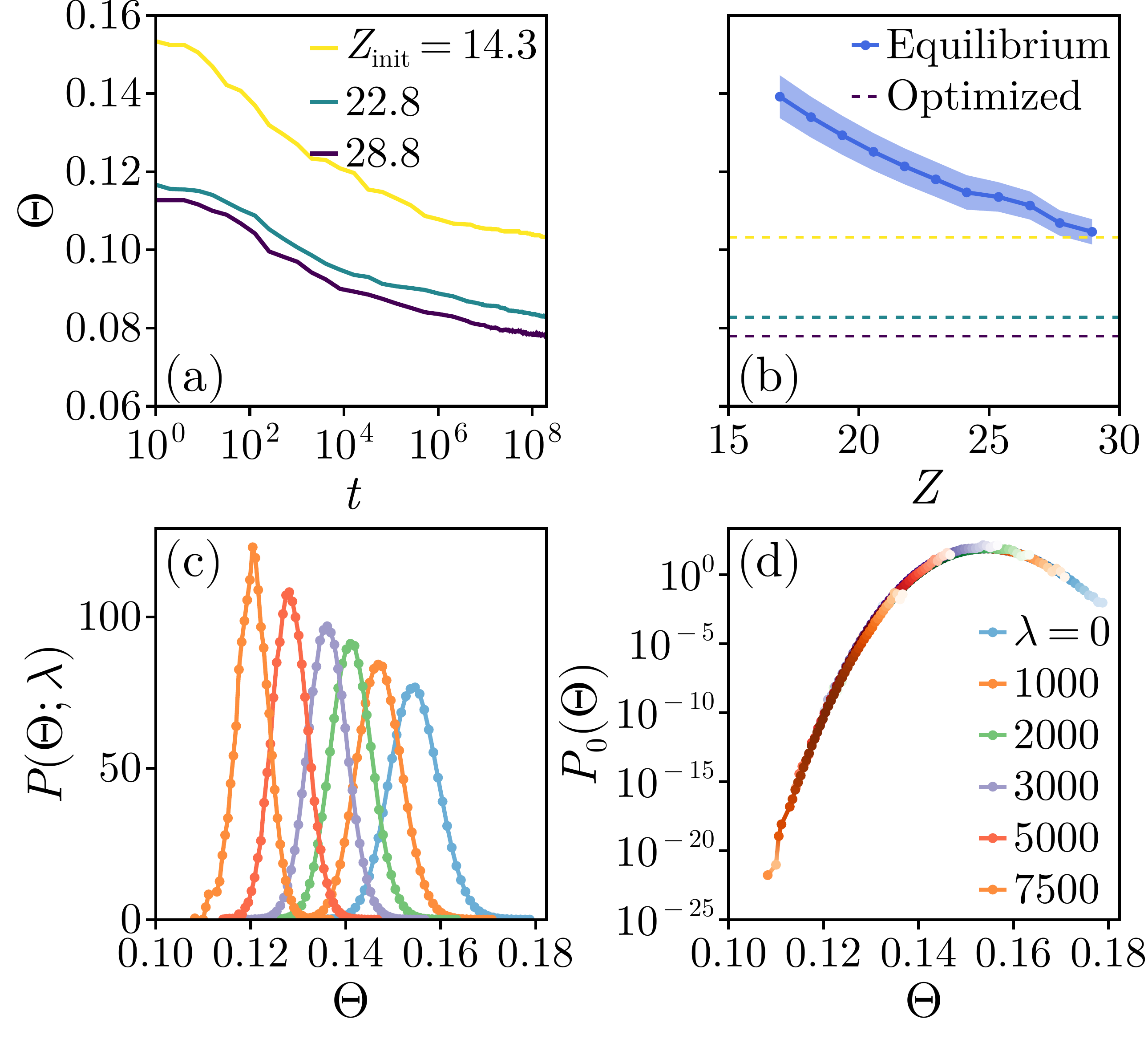} 
\caption{Optimizing the local packing order parameter.
(a) Time evolution of the local order parameter $\Theta$ during a biased Monte Carlo simulations \rev{for $N=200$ particles} with $\lambda = 5 \times 10^5$, starting from equilibrium configurations prepared at different reduced pressure $Z_{\rm init}$.
(b) Comparison of the final values of $\Theta$ measured at time $t=2\times10^8$ (dashed lines) with the equilibrium values already shown in Fig.~\ref{fig:statics}(b).
(c) Probability distributions $P(\Theta; \lambda)$ of the fluctuations of $\Theta$ measured in biased simulations at different values of the bias strength $\lambda$ at $\phi=0.72$. 
(d) Reconstructed equilibrium histogram $P_0(\Theta)$ after reweighting the biased distributions. Configurations with low $\Theta$ values have vanishingly small probability to appear in the equilibrium ensemble, and are non-typical.}  
\label{fig:theta}
\end{figure}

We show in Fig.~\ref{fig:theta}(a) three representative time evolutions of the local order parameter $\Theta$ during a biased Monte Carlo simulation for a system of $N=200$ particles with a bias strength $\lambda = 5 \times 10^5$, using $n=1$.
\rev{The choice of a relatively small system size is dictated by the heavy computational cost of continuously updating the Voronoi tessellation required to evaluate $\Theta$ after each proposed single-particle displacement.}
These trajectories were started, at $t=0$, from equilibrium hard disk configurations at reduced pressure $Z_{\rm init}$. All three trajectories display a significant decrease in the value of $\Theta$ relative to their initial equilibrium values, confirming that the algorithm effectively favors configurations with much better local packing.

In Fig.~\ref{fig:theta}(b), we compare the values of $\Theta$ obtained after $2\times10^8$ Monte Carlo iterations with the biased Hamiltonian to the equilibrium hard disk values already shown in Fig.~\ref{fig:statics}(b). \rev{Since $\Theta$ is a system-averaged quantity, the magnitude of its typical fluctuations scales as $1/\sqrt{N}$. We display them in Fig.~\ref{fig:theta}(b) to provide direct evidence that the optimized configurations lie well beyond the regime explored by thermal fluctuations alone.} Extrapolating the equilibrium data, we estimate that these optimized configurations reach values of $\Theta$ corresponding to equilibrium states at $Z\approx29,\,38,\,40$. The first value is close to our numerical glass transition (with $\tau_\alpha \sim 5 \times 10^9$), while the states originating from the two largest values of $Z_{\text{init}}$ correspond to pressures well beyond our equilibration capabilities, with respective estimates for $\tau_\alpha$ of order $10^{23}$ and $10^{29}$, using the parabolic fit of equilibrium data. Equilibrium systems at such pressure would be representative of ultrastable glasses, and the variations in $\Theta$ that we obtain using the biased Monte Carlo simulations are significant. 

To assess the validity of the algorithm and illustrate its physical behavior, we measure the fluctuations of $\Theta$ in the steady state of biased simulations, using several values of the biasing field $\lambda$. We collect these data in the probability distributions $P(\Theta ; \lambda)$, shown in Fig.~\ref{fig:theta}(c) for the area fraction $\phi=0.72$. As expected, when $\lambda$ increases, the distribution shifts towards lower values of $\Theta$ and becomes slightly narrower.

Since the sampling distribution $P(x)$ is known up to a normalization constant, we can reweight the probability distributions $P(\Theta ; \lambda)$ to express the equilibrium distribution $P_0(\Theta)$ of the fluctuations of $\Theta$ under the Hamiltonian $H_0$, using the relation $P_0(\Theta) \propto P(\Theta; \lambda) \exp\left(\lambda \Theta^2(x)\right)$. This relation is valid for any value of $\lambda$ and holds up to an unknown normalization constant. The reweighting procedure is illustrated in Fig.~\ref{fig:theta}(d) where the six histograms of $\Theta$ are simultaneously used to reconstruct $P_0(\Theta)$, using the unknown normalization to connect the different histograms together~\cite{ferrenberg1989optimized}. The quality of the reconstructed distribution confirms that the reweighting procedure is reliable and that the Monte Carlo biased algorithm correctly samples the biased ensemble, as expected. Furthermore, the equilibrium histogram also demonstrates that the optimized configurations produced by the biased Monte Carlo algorithm are atypical hard disk configurations that would be sampled with a vanishingly small probability using an equilibrium approach. Already for $\lambda=7.5 \times 10^3$, which is far below the values used in Fig.~\ref{fig:theta}(a), the reweighted equilibrium density extends down to values below $10^{-20}$, showing that such biased configurations lie very deep in the tails of the equilibrium distribution, and are indeed vanishingly rare.

\subsection{Stability}

\begin{figure}
\includegraphics[width=\columnwidth]{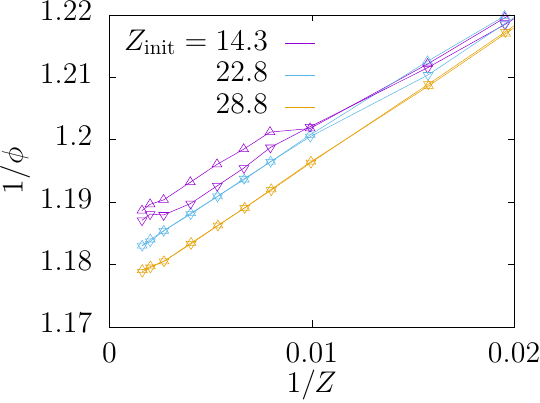} 
\caption{Glass equations of state for glasses prepared using either conventional Monte Carlo methods (upwards triangles) or optimized glasses prepared using biased Monte Carlo simulations (downwards triangles), for three values of $Z_{\rm init}$. Optimized and conventional glasses are equally stable.}
\label{fig:EOSlambda}
\end{figure}

\rev{To test the thermodynamic stability} of the optimized glass configurations with enhanced local packing obtained using the biased Monte Carlo simulations, we repeat the measurement of the glass equations of state, as described in Sec.~\ref{sec:glassEOS} above. 

We quantitatively compare two families of approaches. On the one hand, we can use the equilibrium (non-hyperuniform) hard disk configurations at $Z_{\rm init}$ to produce conventional glasses obtained by rapid compressions to large pressure, as already done in Sec.~\ref{sec:glassEOS}. On the other hand, we can also compress the optimized configurations to large pressures, and measure their glass equations of state during the same decompression protocols.  

Since we have three values of $Z_{\rm init}$ and two families of glasses (conventional and locally optimized), we report six glass equations of state in Fig.~\ref{fig:EOSlambda}. Each color represents a given value of $Z_{\rm init}$ while different symbols are used to distinguish conventional from optimized glasses. Clearly, the data in Fig.~\ref{fig:EOSlambda} are grouped by colors. This implies that glass stability is uniquely controlled by $Z_{\rm init}$, independently of the value of $\Theta$. In other words, optimizing the local order parameter $\Theta$ to a degree that can only be obtained in very stable glasses has not influenced the initial stability at all. The logical conclusion is that the local order parameter $\Theta$ takes small values in highly stable glasses, but glasses with optimized $\Theta$ are not necessarily stable. Glass stability and $\Theta$ are correlated physical properties, but the local order parameter $\Theta$ is not causally responsible for enhanced stability. 

\rev{We have also applied our biased Monte Carlo algorithm to configurations that are already dynamically arrested deep in the glass phase. We found that it is also possible to decrease significantly the value of $\Theta$ in these glasses, although this becomes increasingly difficult as the pressure increases. This is expected, as particle motion is no longer allowed in the limit $Z \to \infty$. More importantly, we observed that the packing fraction of these glasses is not evolving with $\Theta$. This directly shows that the glass equation of state, and therefore the thermodynamic stability, is not improved by the optimization protocol. This suggests that performing the local optimization after the glass has been formed does not allow the system to access denser, more stable regions of the landscape.}

Earlier observations of stable glasses with low $\Theta$ values resulted from two ingredients~\cite{fan2024ideal}: a coupled position-diameter dynamics and non-equilibrium rules that optimize $\Theta$. Our interpretation is that it is the coupled position-diameter dynamics that is causally responsible for enhanced stability, rather than the enhanced local order. We discuss this conclusion further in the discussion part in Sec.~\ref{sec:conclusion}.

\section{Discussion and perspectives}

\label{sec:conclusion}

The search for glasses that lie very deep in their potential energy landscape is motivated by both fundamental and practical aspects~\cite{berthier2016facets,ediger2017highly,ninarello2026,royall2018race}. At the fundamental level, low-energy glassy states are useful to address key questions regarding the underlying nature of the glass transition, such as the putative existence of an equilibrium ideal glass phase~\cite{berthier2011theoretical}. When it comes to practical applications, being close to equilibrium is not needed, provided the resulting glassy materials have interesting physical properties~\cite{ediger2017highly,rodriguez2022ultrastable}. For these reasons, exploring multiple routes to design glasses is an interesting endeavor~\cite{berthier2026designing,ninarello2026}.  

A recent series of papers explored the idea that optimizing a physical property that is not directly the potential energy may nevertheless lead to glassy states with enhanced stability. These approaches share a common strategy to optimize a given physical property, through which configurations evolve using both positions and diameters as dynamic degrees of freedom. This non-trivial coupled dynamics was implemented within minimization processes that optimize hyperuniformity at large length scales~\cite{yana2021towards,dale2022hyperuniform,wang2025hyperuniform}, or the fluctuations of the Virial stress~\cite{leoni2025generating}, or a local packing order parameter~\cite{fan2024ideal}, and of course the potential energy itself~\cite{varda2002transient,brito2018theory,kapteijns2019fast,bolton2026ideal}. \rev{In all these algorithms, the particle diameters are dynamically altered, and the particle size distribution is not conserved.}

Our main result is the demonstration that the sole optimization of either hyperuniformity or the local packing order parameter in the absence of the coupled position-diameter dynamics does not improve the stability of glasses. We conclude that more stable glasses have improved physical properties, but improved physical properties do not necessarily imply a greater stability.

The most plausible interpretation is that the physical properties targeted in these works are statistically strongly correlated with the potential energy of the system. When preparing glasses with enhanced stability, all these properties are then evolving smoothly together. However, to improve stability, the system has to evolve across large barriers in a complex energy landscape. In the algorithms used in Refs.~\cite{yana2021towards,dale2022hyperuniform,wang2025hyperuniform,fan2024ideal,leoni2025generating,varda2002transient,brito2018theory,kapteijns2019fast,bolton2026ideal}, it is very likely that it is the coupled diameter-position microscopic dynamics that helps the system efficiently cross these barriers, while the choice of the physical property to be optimized is a sub-dominant ingredient. 

Additional support for this conclusion comes from a series of recent studies of the effect of diameter dynamics in the simulations of model glass-formers, which we now discuss. The first and most obvious observation is that the dramatic equilibration speedup offered by the swap Monte Carlo algorithm results precisely from the dynamic coupling between diameters and positions during the course of a simulation~\cite{ninarello2017models,berthier2019efficient,szamel2018theory,ikeda2017mean}.

The connection between swap Monte Carlo and the algorithms employed in Refs.~\cite{yana2021towards,dale2022hyperuniform,wang2025hyperuniform,fan2024ideal,leoni2025generating} is relatively direct. Consider, as in those works, models with a continuous distribution of particle sizes. When two particles, 1 and 2, are swapped, $1 \leftrightarrow 2$, one can either exchange the positions of the two particles, ${\mathbf r}_1 \leftrightarrow {\mathbf r}_2$, or leave the positions unaffected but exchange their diameters, $\sigma_1 \leftrightarrow \sigma_2$. In the latter view, adopted in Refs.~\cite{berthier2016equilibrium,ninarello2017models} and most subsequent papers, the diameter of a given particle is promoted to a dynamic variable whose evolution is coupled to that of its position. This dynamic coupling is directly responsible for the efficiency of the swap Monte Carlo algorithm to equilibrate supercooled liquids at much lower temperatures than conventional dynamics. 

While a swap Monte Carlo move leaves the particle size distribution unaffected, one can promote the swap dynamics to a semi-grand canonical version, where each diameter can now evolve freely and continuously via exchange with a diameter reservoir. If constraints are correctly added to the corresponding chemical potential, it is possible to maintain a fixed particle size distribution and simulate equilibrium dynamics~\cite{berthier2019efficient}. It was demonstrated numerically that the original swap Monte Carlo and such a semi-grand canonical version are dynamically equivalent~\cite{berthier2019efficient}. When the particle size distribution was insufficiently constrained, however, it was found that in the presence of thermal fluctuations the system modifies its particle size distribution to more easily crystallize or fractionate~\cite{berthier2019efficient}. 

One solution to prevent such ordering is to remove thermal fluctuations entirely, as done for instance in Refs.~\cite{brito2018theory,kapteijns2019fast,varda2002transient}. Here, the potential energy of the system was directly minimized by following a gradient descent where both positions and diameters could evolve. In a traditional energy minimization, only the positions are evolving. The addition of diameters as dynamic degrees of freedom allows the system to reach much deeper energy states, and this leads to more stable glasses~\cite{kapteijns2019fast,varda2002transient}. It was recently shown that equivalent results are obtained by performing a zero-temperature swap Monte Carlo simulation~\cite{nishikawa2025irreversible} (or an infinite pressure compression for hard particles~\cite{berthier2016equilibrium,ghimenti2024irreversible,berthier2024monte}), showing that gradient descent with evolving diameters is fully analogous to zero-temperature swap Monte Carlo dynamics.  

The algorithms in Refs.~\cite{yana2021towards,dale2022hyperuniform,wang2025hyperuniform,fan2024ideal,leoni2025generating} can be regarded as slight modifications of the above efforts, where the quantity to be minimized is no longer the potential energy but a different physical quantity. However, for strongly correlated quantities, minimizing with respect to one of them is not very different from minimizing with respect to another. \rev{The deep connections between computational strategies described above suggest} that all these algorithms presumably explore similar paths in the glassy landscape, and, most importantly, they can easily cross barriers due to the modified dynamics that couples diameters and positions.

This interpretation immediately raises several important questions, which would be worth analyzing in the future. First, one could question whether the new freedom offered by the absence of constraints on the particle size distribution plays a role in improving the glass properties. One potential danger is that the system uses that freedom to develop some kind of (possibly unwanted) local ordering~\cite{berthier2019efficient}. This is possibly mitigated by the absence of thermal fluctuations, but it could be that improved stability also stems, in part, from some increase in local order. As a second line of questions, it would be interesting to benchmark various algorithms based on different physical quantities against one another. One could discover that some quantities are better suited than others and lead to better stability or distinct glassy states. Instead, it could also be that all of them are essentially equivalent. This is an important task for future work. However, since each algorithm produces a distinct glassy material, this quest also underlies the need to develop objective metrics to quantitatively compare the different outcomes of different design rules.

All these questions are broadly applicable to experimental studies as well, even though the algorithms discussed in the present work are not directly related to any experimental realization. In experimental work, glass stability is dramatically enhanced when physical vapor deposition is used~\cite{swallen2007organic,ediger2017highly,rodriguez2022ultrastable}. In that case, the structure of the glass is optimized compared to conventional bulk materials (for instance some ultrastable glasses display structural anisotropy~\cite{dalal2015tunable}), but the dynamic process used to create them is also very different from the bulk dynamics. In that case also, it is generally believed that ultrastability is controlled by the dynamic process (the large decoupling between surface and bulk dynamics in amorphous films~\cite{zhu2011surface,daley2012comparing,berthier2017origin}) rather than the increased structural anisotropy.  For experiments as well, the determination of objective metrics to compare different glassy materials is a relevant task for future research. 
  
\section*{Data Availability}

\rev{The data that support the findings of this study are openly available~\cite{zenododataset}}.

\acknowledgments

We thank D. Coslovich, M. Ediger, R. Jack, J. Kurchan and G. Tarjus for useful exchanges about this work. L.B. acknowledges the support of the French Agence Nationale de la Recherche (ANR), under grants ANR-20-CE30-0031 (project THEMA) and ANR-24-CE30-0442 (project GLASSGO). 

\bibliography{main.bib}

\end{document}